\documentclass{amsart}

\usepackage{latexsym}
\usepackage{amsfonts}
\begin{document}

\newcommand{\bm}[1]{\mbox{\boldmath $\bf#1$}}

\def\b{\begin{equation}}
\def\e{\end{equation}}

\def\ba{\begin{eqnarray}}
\def\ea{\end{eqnarray}}
\newtheorem{theorem}{Theorem}[section]
\newtheorem{lemma}[theorem]{Lemma}
\newtheorem{proposition}[theorem]{Proposition}
\newtheorem{remark}[theorem]{Remark}

\theoremstyle{definition}
\newtheorem{definition}[theorem]{Definition}

\theoremstyle{remark}
\newtheorem{conjecture}[theorem]{Conjecture}

\theoremstyle{remark}

\title{B\"acklund transformations for fourth Painlev\'e hierarchies}
\author{Pilar R.~Gordoa}
\thanks{{\em 2000 Mathematics Subject Classification(s):} 
Primary 34M55; Secondary 33E17, 35Q51, 35Q58}
\thanks{Pilar R.~Gordoa,
Area de Matem\'atica Aplicada, ESCET, Universidad Rey Juan Carlos, 
C/~Tulip\'an s/n, 28933 M\'ostoles, Madrid, Spain.
({\em E-mail address}: prg@sonia.usal.es)}
\author{Nalini Joshi}
\thanks{Nalini Joshi, School of Mathematics and Statistics, University of Sydney,
NSW2006 Sydney, Australia. ({\em E-mail address}: Nalini.Joshi@maths.usyd.edu.au)}
\author{Andrew Pickering}
\thanks{Andrew Pickering, Departamento de Matem\'aticas, Universidad de Salamanca,
Plaza de la Merced 1, 37008 Salamanca, Spain. ({\em E-mail address}: andrew@gugu.usal.es)}
\commby{XXX}

\begin{abstract}
B\"acklund transformations (BTs) for ordinary differential equations 
(ODEs), and in particular for hierarchies of ODEs, are a topic of 
great current interest. Here we give an improved method of constructing 
BTs for hierarchies of ODEs. This approach is then applied to fourth 
Painlev\'e ($P_{IV}$) hierarchies recently found by the same authors 
[{\em Publ. Res. Inst. Math. Sci. (Kyoto)} {\bf 37} 327--347 (2001)]. 
We show how the known pattern of BTs for $P_{IV}$ can be extended
to our $P_{IV}$ hierarchies. Remarkably, the BTs required to do this 
are precisely the Miura maps of the dispersive water wave hierarchy.
We also obtain the important result that the fourth Painlev\'e equation 
has only one nontrivial fundamental BT, and not two such as is frequently
stated.
\end{abstract}

\maketitle
\section{Introduction}

A classical problem, dating from the end of the nineteenth century,
is that of seeking new transcendental functions defined by ordinary
differential equations (ODEs). This motivated the classification of
ODEs having what is today referred to as the Painlev\'e property,
i.e.\ having their general solution free of movable branched 
singularities. In particular, it led to the discovery of the six
Painlev\'e equations \cite{P00,P02,G10,Ince}, which did indeed define 
new transcendental functions.

The six Painlev\'e equations are of course second order ODEs. However
the classification program embarked upon by Painlev\'e and co-workers 
foresaw, once second order ODEs had been dealt with, a classification
of third order ODEs, then of fourth order ODEs, and so on. Thus Chazy
\cite{C11} and Garnier \cite{G12} studied certain classes of third 
order ODEs, although no new transcendent was discovered at third order.
Restricted classes of third order ODEs were also later considered by 
Exton \cite{E71} and Martynov \cite{M85a,M85b}, unfortunately with the 
same result. It should be remarked that the difficulties of classification 
increase with the order of the equations studied; for example, at second 
order movable essential singularities may arise \cite{P02}, whereas at 
third order movable natural boundaries may occur \cite{C11}. At fourth
order, even the classification of dominant terms for the polynomial case 
was left incomplete \cite{B64}.

Thus, some 20-25 years ago, the search for higher order ODEs defining new
transcendental functions was in need of a new insight in order to catalyse 
research in this area. This impetus came in the form of the discovery by 
Ablowitz and Segur \cite{AS77} of a connection between completely integrable 
partial differential equations (PDEs) and ODEs having the Painlev\'e 
property. This discovery not only made a remarkable connection between 
modern research and mathematics at the turn of the last century, but in 
establishing a link between integrability and the analytical properties
of solutions, mirrored the prize-winning work of Kowalevski on the motion 
of a rigid body about a fixed point \cite{K89a,K89b}. It was Airault who,
exploiting the fact that, for example, sitting above the Korteweg-de Vries
(KdV) equation and the modified KdV (mKdV) equation are their respective
hierarchies, first realised the next step of using higher order integrable
PDEs to derive higher order ODEs with the Painlev\'e property. In fact she
derived a whole hierarchy of ODEs, the second Painlev\'e ($P_{II}$) 
hierarchy, by similarity reduction of the KdV\!/mKdV hierarchies \cite{Air}.

However Airault also made another important step: she obtained B\"acklund 
transformations (BTs) for every member of the $P_{II}$ hierarchy. A BT is 
a mapping between solutions of ODEs, involving naturally some identification 
between the parameters appearing as coefficients in the ODEs; in the case of 
BTs between solutions of the same ODE, this identification between parameters 
translates as changes in paramater values. BTs for the Painlev\'e equations 
had previously been studied in the Soviet literature; a comprehensive list of 
references can be found in \cite{FA82}, and a recent review in \cite{GLS}. 
Today BTs are universally recognised as an important property of integrable
nonlinear ODEs, and there is much interest in their derivation, especially 
within the context of hierarchies of ODEs. The aim of the present paper is
to explore BTs for fourth Painlev\'e ($P_{IV}$) hierarchies.

Due therefore to the work of Ablowitz and Segur, over the last quarter
century, the study of higher order analogues of the Painlev\'e equations,
and of their properties, has been informed by knowledge of the connection 
with completely integrable PDEs; here we refer for example to the work of 
Mu\u gan and Jrad \cite{MJ99,MJ02,MJp}, and Cosgrove \cite{C00}. The present 
authors have also exploited this connection \cite{GP99a,GP99b,GP00,GJP01} in 
the development of their own method \cite{GP99b} of deriving  (amongst other 
things) hierarchies of higher order Painlev\'e equations together with 
associated underlying linear problems. Here we extend this connection
still further: we find that certain features of such ODEs are directly 
related to the underlying structures of associated completely integrable 
PDEs. That is, when seeking to extend to a fourth Painlev\'e hierarchy 
\cite{GJP01} the pattern of BTs already known for the first member 
($P_{IV}$), we find that the answer lies in the Miura transformations for 
the associated PDE hierarchy.

The layout of the paper is as follows. We introduce our
$P_{IV}$ hierarchies in Section Two. In Section Three we give an improved method, based 
on the Painlev\'e truncation process for PDEs, of deriving auto-BTs and 
special integrals for hierarchies of ODEs, and as an example we apply this 
approach to the $P_{II}$ hierarchy of Airault. In Section Four we use this 
method to derive auto-BTs and special integrals for two of the $P_{IV}$ 
hierarchies derived in Section Two. In Section Five we identify to which
BTs of $P_{IV}$ these BTs correspond. In Section Six we seek further BTs
in order to extend the known pattern of BTs for $P_{IV}$ to corresponding
hierarchies. Remarkably, it turns out that the BTs required to do this are 
precisely the known Miura maps for the associated PDE (dispersive water 
wave) hierarchy. In Section Seven we consider a mapping between our
hierarchies which allows us to further relate the BTs derived: an important
consequence of this is the result that $P_{IV}$ has only one 
nontrivial fundamental BT. Section Eight is devoted to conclusions.

\section{Sequences of fourth Painlev\'e hierarchies}

In our recent paper \cite{GJP01} we derived, along with associated linear
problems, the sequence of coupled ODEs in ${\bf u}=(u,v)^T$,
\begin{equation}
{\mathcal R}^n{\bf u}_x+\sum_{i=0}^{n-1}c_i{\mathcal R}^i{\bf u}_x
+g_{n-1}{\mathcal R}^2\left(\begin{array}{c} 1 \\ 0 \end{array}\right)+g_n
{\mathcal R}\left(\begin{array}{c} 1 \\ 0 \end{array}\right) +g_{n+1} \left(\begin{array}{c} 
1 \\ 0 \end{array}\right) = \left(\begin{array}{c} 0 \\ 0 \end{array}\right),
\label{gdww}
\end{equation}
where $c_0,\ldots,c_{n-1}$, $g_{n-1}$, $g_n$ and $g_{n+1}$,
are arbitrary constants, and ${\mathcal R}$ is the recursion operator of the 
dispersive water wave (DWW) hierarchy \cite{B75}---\cite{sachs} 
($\partial_x=\partial /\partial x=d/dx$ in our ODE case (\ref{gdww})),
\begin{equation}
\label{rec}
{\mathcal R} = \frac{1}{2}\left(\begin{array}{cc}\partial_x u \partial_x^{-1} 
           -\partial_x & 2 \\
           2v+v_x\partial_x^{-1} & u+\partial_x \end{array}\right).
\end{equation}
In what follows we will consider the case which corresponds to a generalized
$P_{IV}$ hierarchy, i.e.\ $g_{n-1}=0$ and $g_n\neq0$ \cite{GJP01}. We can then 
assume, using a shift on $u$, that $g_{n+1}=0$ (note that previously we have 
used such a shift to set $c_{n-1}=0$, but here we prefer to remove $g_{n+1}$).
Further, using a shift on $x$, we can set $c_0=0$. Thus, without any loss of
generality, we can assume that our generalized $P_{IV}$ hierarchy is of the
form
\begin{equation}
{\mathcal R}^n{\bf u}_x+\sum_{i=1}^{n-1}h_i{\mathcal R}^i{\bf u}_x
+g_n {\mathcal R}\left(\begin{array}{c} 1 \\ 0 \end{array}\right) 
= \left(\begin{array}{c} 0 \\ 0 \end{array}\right)
\label{gdw1}
\end{equation}
for some constants $h_1,\ldots,h_{n-1}$ and $g_n(\neq0)$. We note in passing
that the second nontrivial member of our hierarchy ($n=2$) is of interest for
the problems that its singularity analysis presents; this was the subject of
our paper \cite{GJP02}.

The hierarchy (\ref{gdw1}) can also be written in the alternative form
\b
B_2{\bf K}_n[{\bf u}]=0,
\label{gdw2}
\e
where
\b
{\bf K}_n[{\bf u}]={\bf L}_n[{\bf u}]+\sum_{i=1}^{n-1}h_i{\bf L}_i[{\bf u}]
+g_n\left(\begin{array}{c} 0 \\ x \end{array}\right),
\e
$B_2$ is one of the three Hamiltonian operators of the DWW hierarchy,
\b
B_2 = \frac{1}{2}\left(\begin{array}{cc} 2\partial_x &
      \partial_x u-\partial_x^2 \\
      u\partial_x+\partial_x^2 & 
      v\partial_x+\partial_x v \end{array}\right),
\label{B2}
\e
and each ${\bf L}_i[{\bf u}]$ is the variational derivative of the 
Hamiltonian density corresponding to the operator $B_2$ for the 
$t_i$-flow of the DWW hierarchy, 
${\bf u}_{t_i}={\mathcal R}^i{\bf u}_x=B_2{\bf L}_i[{\bf u}]$.

Here we have used the fact that $\mathcal R=B_2B_1^{-1}$, where
\b
B_1 = \left(\begin{array}{cc} 0 & \partial_x \\ 
      \partial_x & 0 \end{array}\right)
\e
is another of the Hamiltonian operators of the DWW hierarchy.
We note also the recursion relation 
$B_1{\bf L}_{i+1}[{\bf u}]=B_2{\bf L}_i[{\bf u}]$, and that
${\bf L}_0[{\bf u}]=(0,2)^T$, ${\bf L}_1[{\bf u}]=(v,u)^T$.

We now consider the construction of hierarchies equivalent to
(\ref{gdw2}). We will also see how a reduction of order of our
system (\ref{gdw2}) can be effected using the Hamiltonian
structures of the DWW hierarchy. 
We begin by recalling the Miura maps of the DWW hierarchy, as given 
by Kupershmidt \cite{Kup}. The first Miura map is given by 
${\bf u}={\bf F}[{\bf U}]$, where ${\bf U}=(U,V)^T$ and
\b
{\bf F}[{\bf U}]=\left(\begin{array}{c} U \\ UV-V^2+V_x \end{array}\right).
\e
The two second Miura maps are given by ${\bf U}={\bf \Phi}[\bm{\phi}]$,
where $\bm{\phi}=(\phi,p)^T$ and
\b
{\bf \Phi}[\bm{\phi}]=\left(\begin{array}{c} \phi+2p \\ p \end{array}\right),
\label{mm2a}
\e
and ${\bf U}={\bf \Psi}[\bm{\psi}]$, where $\bm{\psi}=(\psi,s)^T$ and
\b
{\bf \Psi}[\bm{\psi}]=\left(\begin{array}{c} \psi-2s \\ -s \end{array}\right).
\label{mm2b}
\e
That is, we have the following two sequences of Miura transformations:
\b
\left(\begin{array}{c} u \\ v \end{array}\right)\ 
\stackrel{\vector(1,0){50}}{\bf F}
\left(\begin{array}{c} U \\ V \end{array}\right)\ 
\begin{array}{c}
{\stackrel{\vector(1,0){50}}{\bf \Phi}} 
\left(\begin{array}{c} \phi \\ p \end{array}\right)\\
{\stackrel{\vector(1,0){50}}{\bf \Psi}}
\left(\begin{array}{c} \psi \\ s \end{array}\right)
\end{array}
\label{seq}
\e

We now consider the first of these sequences. Using the fact that for the 
Miura map $\bf F$ we have 
\b
\begin{array}{l|rc}
B_2 & & = {\bf F}'[{\bf U}] B  \left({\bf F}'[{\bf U}]\right)^{\dagger}, \\
 & {_{{\bf u}={\bf F}[{\bf U}]}}
\end{array}
\label{facF}
\e
where $B$ is the Hamiltonian operator of the modified DWW system,
\b
B = \frac{1}{2}\left(\begin{array}{cc} 2\partial_x & \partial_x
\\ \partial_x & 0 \end{array}\right),
\e
${\bf F}'[{\bf U}]$ is the Fr\'echet derivative of the Miura map and
$({\bf F}'[{\bf U}])^{\dagger}$ is its adjoint, we obtain, in the same way
as in the PDE case, the modified version of (\ref{gdw2}),
\b
B\left({\bf F}'[{\bf U}]\right)^{\dagger}{\bf K}_n[{\bf F}[{\bf U}]]=0.
\e
This we can then integrate to obtain
\b
\left({\bf F}'[{\bf U}]\right)^{\dagger}{\bf K}_n[{\bf F}[{\bf U}]]
+(c_n,d_n)^T=0,
\label{mod1}
\e
for two arbitrary constants $c_n$ and $d_n$. In fact this last is 
equivalent to an integrated version of (\ref{gdw2}) under the BT
\ba
{\bf u}-{\bf F}[{\bf U}] & = & 0, 
\label{bt1a} \\
\left({\bf F}'[{\bf U}]\right)^{\dagger}{\bf K}_n[{\bf u}]
+(c_n,d_n)^T & = & 0.
\label{bt1b}
\ea
It is this construction of a BT between an integrated modified hierarchy
and an integrated version of our original hierarchy that lies behind the
first integrals of our $P_{IV}$ hierarchy given in \cite{GJP01}; this
approach is described in more detail in \cite{AP02}. The integrated form
of (\ref{gdw2}) obtained from (\ref{bt1a}), (\ref{bt1b}) is
\ba
L_{n,x} & = & 2K_n+uL_n+(g_n-2\alpha_n), \label{si1} \\
K_{n,x} & = & \frac{(K_n+\frac{1}{2}g_n-\alpha_n)^2-\frac{1}{4}\beta_n^2}
{L_n}-vL_n,\label{si2}
\ea
where ${\bf K}_n=(K_n,L_n)^T$, and where we have set 
$2c_n+d_n=g_n-2\alpha_n$
and $d_n^2=\beta_n^2$.

In the same way, since under the composition ${\bf H}={\bf F}\circ{\bf \Phi}$ 
we have analogously to (\ref{facF})
\b
\begin{array}{l|rc}
B_2 & & = {\bf H}'[\bm{\phi}] C  \left({\bf H}'[\bm{\phi}]\right)^{\dagger} \\
 & {_{{\bf u}={\bf H}[\bm{\phi}]}}
\end{array}
\label{facH}
\e
with
\b
C = \frac{1}{2}\left(\begin{array}{cc} -2\partial_x & \partial_x
\\ \partial_x & 0 \end{array}\right),
\e
we obtain the integrated second modified hierarchy, 
\b
\left({\bf H}'[\bm{\phi}]\right)^{\dagger}{\bf K}_n[{\bf H}[\bm{\phi}]]
+(e_n,f_n)^T=0.
\label{mod1a}
\e
It is easy to see that the constants of integration in (\ref{mod1}) and
(\ref{mod1a}) are related by $c_n=e_n$ and $d_n=f_n-2e_n$.

For our second sequence of Miura transformations we have the composition
${\bf I}={\bf F}\circ{\bf \Psi}$ and, corresponding to (\ref{facH}),
\b
\begin{array}{l|rc}
B_2 & & = {\bf I}'[\bm{\psi}] D  \left({\bf I}'[\bm{\psi}]\right)^{\dagger} \\
 & {_{{\bf u}={\bf I}[\bm{\psi}]}}
\end{array}
\label{facI}
\e
with
\b
D = \frac{1}{2}\left(\begin{array}{cc} -2\partial_x & -\partial_x
\\ -\partial_x & 0 \end{array}\right).
\e
Thus we obtain the alternative integrated second modified hierarchy, 
\b
\left({\bf I}'[\bm{\psi}]\right)^{\dagger}{\bf K}_n[{\bf I}[\bm{\psi}]]
+(l_n,m_n)^T=0,
\label{mod1b}
\e
with constants of integration related to those of (\ref{mod1}) by
$c_n=l_n$ and $d_n=-m_n-2l_n$.

The hierarchies (\ref{si1})---(\ref{si2}), (\ref{mod1}), (\ref{mod1a}) 
and (\ref{mod1b}) are all $P_{IV}$ hierarchies. In order to show this,
let us consider the case $n=1$ of these hierarchies. We have $K_1=v$
and $L_1=u+g_1x$, and so our system (\ref{si1})---(\ref{si2}) reads
\ba
u_x & = & 2v+u(u+g_1x)-2\alpha_1, \\
v_x & = & \frac{(v+\frac{1}{2}g_1-\alpha_1)^2-\frac{1}{4}\beta_1^2}{u+g_1x}
         -v(u+g_1x);
\ea
eliminating $v$ and setting $u=\pm y-g_1x$ yields the fourth Painlev\'e equation,
\b
y_{xx}=\frac{1}{2}\frac{y_x^2}{y}+\frac{3}{2}y^3\mp 2g_1xy^2+2\left(\frac{1}{4}
g_1^2x^2-\alpha_1\right)y-\frac{1}{2}\frac{\beta_1^2}{y}.
\label{p4y}
\e
The system (\ref{mod1}) reads
\ba
V_x+2UV-V^2+g_1xV+c_1 & = & 0, \label{UV1} \\
U_x+2UV-U^2-g_1(U-2V)x+g_1-d_1 & = & 0. \label{UV2}
\ea
Elimination of $V$ and setting $U=\pm y-g_1x$, 
$2c_1+d_1=g_1-2\alpha_1$ and $d_1^2=\beta_1^2$ yields (\ref{p4y}).
This follows immediately from the fact that in the Miura map $U=u$. However,
eliminating $U$ and setting $V=\pm w$ also yields the fourth Painlev\'e 
equation,
\b
w_{xx}=\frac{1}{2}\frac{w_x^2}{w}+\frac{3}{2}w^3\pm 2g_1xw^2+2\left[\frac{1}{4}
g_1^2x^2-\frac{1}{2}(c_1+2d_1-g_1)\right]w-\frac{1}{2}\frac{c_1^2}{w}.
\label{p4w}
\e
Thus for $n=1$, both independent variables of the first modification define
versions of $P_{IV}$. We will return to the relationship between these two 
copies of $P_{IV}$ later.

Our first second modification (\ref{mod1a}), for $n=1$, reads
\ba
p_x +2\phi p+3p^2 +g_1 x p +e_1 & = & 0, \label{m21a} \\
\phi_x -6\phi p -6p^2-\phi^2-g_1x(\phi+2p)+g_1-f_1 & = & 0; \label{m21b}
\ea
eliminating $\phi$ yields
\b
p_{xx}=\frac{1}{2}\frac{p_x^2}{p}+\frac{3}{2}p^3+2g_1xp^2+2\left[\frac{1}{4}
g_1^2x^2-\frac{1}{2}(2f_1-3e_1-g_1)\right]p-\frac{1}{2}\frac{e_1^2}{p},
\label{p4p}
\e
i.e.\ the fourth Painlev\'e equation. Noting that in the Miura map $V=p$, we
see that this last is equivalent to (\ref{p4w}), for the upper choice of sign,
with the identification $w=p$, $c_1=e_1$ and $d_1=f_1-2e_1$.

Our second second modification (\ref{mod1b}), for $n=1$, reads 
\ba
s_x +2\psi s-3s^2 +g_1 x s -l_1 & = & 0, \label{n21a} \\
\psi_x +6\psi s -6s^2-\psi^2-g_1x(\psi-2s)+g_1+m_1 & = & 0, \label{n21b}
\ea
equivalent to (\ref{m21a}), (\ref{m21b}) under 
$(\phi,p,e_1,f_1)\rightarrow (\psi,-s,l_1,-m_1)$.
Elimination of $\psi$ gives
\b
s_{xx}=\frac{1}{2}\frac{s_x^2}{s}+\frac{3}{2}s^3-2g_1xs^2+2\left[\frac{1}{4}
g_1^2x^2+\frac{1}{2}(2m_1+3l_1+g_1)\right]s-\frac{1}{2}\frac{l_1^2}{s},
\label{p4s}
\e
another version of $P_{IV}$ equivalent to (\ref{p4w}) for the lower choice
of sign, with $w=s$, $c_1=l_1$ and $d_1=-m_1-2l_1$.

Thus we see that the hierarchies (\ref{si1})---(\ref{si2}), (\ref{mod1}), 
(\ref{mod1a}) and (\ref{mod1b}) define sequences of $P_{IV}$ hierarchies,
as in (\ref{seq}). In Section Four we will derive BTs for the hierarchies 
(\ref{mod1a}) and (\ref{mod1b}), and show in Section Six how the known 
structure of BTs for $P_{IV}$ can be replicated for $P_{IV}$ hierarchies, 
using the Miura transformations given above.

Before turning to the derivation of BTs and special integrals for $P_{IV}$
hierarchies, however, we present first of all an improved method of deriving 
BTs for hierarchies of ODEs. As a simple but illuminating example, we apply
this to the $P_{II}$ hierarchy.

\section{B\"acklund transformations for the second Painlev\'e hierarchy}

We take the $P_{II}$ hierarchy in the form
\b
(\partial_x+2Y)\left(M_n[Y_x-Y^2]-\frac{1}{2}x\right)+\frac{1}{2}-\lambda_n=0,
\label{p2h}
\e
where $\lambda_n$ are arbitrary parameters, and the sequence $M_n$ satisfies 
the Lenard recursion relation \cite{Lax} 
$\partial_xM_{n+1}[W]=(\partial_x^3+4W\partial_x+2W_x)M_n[W]$,
with $M_0=1/2$, $M_1[W]=W$. In order to construct a BT for this hierarchy,
we consider adapting the approach developed by Weiss for PDEs \cite{W83},
and seek a ``truncated Painlev\'e expansion''
\b
Y=-\frac{\varphi_x}{\varphi}+\tilde Y,
\label{Ytr}
\e
where
\b
\tilde Y=\frac{1}{2}\frac{\varphi_{xx}}{\varphi_x}.
\label{Ytr1}
\e
For $Y$ defined by (\ref{Ytr}), we find that
\b
Y_x-Y^2=\tilde Y_x-\tilde Y^2-\frac{\varphi_{xx}}{\varphi}
+2\frac{\varphi_x}{\varphi}\tilde Y=\tilde Y_x-\tilde Y^2,
\e
where in order to obtain the last equality we have used (\ref{Ytr1}). 
That is, the quantity $Y_x-Y^2$ is invariant under the mapping 
(\ref{Ytr}), (\ref{Ytr1}). Thus substituting (\ref{Ytr}) into (\ref{p2h}) 
yields
\b
\left(\partial_x+2\tilde Y-2\frac{\varphi_x}{\varphi}\right)
\left(M_n[\tilde Y_x-\tilde Y^2]-\frac{1}{2}x\right)+\frac{1}{2}-\lambda_n=0,
\label{wombat}
\e
Assuming now that $\tilde Y$ also satisfies the corresponding member of the
$P_{II}$ hierarchy, but now for parameter value $\tilde \lambda_n$, i.e.
\b
(\partial_x+2\tilde Y)\left(M_n[\tilde Y_x-\tilde Y^2]-\frac{1}{2}x\right)
+\frac{1}{2}-\tilde \lambda_n=0,
\label{p2t}
\e
we obtain using (\ref{wombat}) and this last,
\b
\frac{\varphi_x}{\varphi}=\frac{\tilde\lambda_n-\lambda_n}
{2M_n[\tilde Y_x-\tilde Y^2]-x}.
\label{bilby}
\e
But (\ref{bilby}) must be compatible with (\ref{Ytr1}), or equivalently 
with the Riccati equation
\b
\left(\frac{\varphi_x}{\varphi}\right)_x+
\left(\frac{\varphi_x}{\varphi}\right)^2
-2\tilde Y \left(\frac{\varphi_x}{\varphi}\right)=0;
\label{ric}
\e
substituting (\ref{bilby}) in (\ref{ric}) gives
\b
(\partial_x+2\tilde Y)\left(M_n[\tilde Y_x-\tilde Y^2]-\frac{1}{2}x\right)
+\frac{1}{2}(\lambda_n-\tilde\lambda_n)=0,
\e
and so comparing with (\ref{p2t}) we see that this compatibility requires
\b
\lambda_n+\tilde \lambda_n=1.
\label{p2s}
\e
Thus we obtain Airault's BT \cite{Air}
\b
Y=\tilde Y+\frac{\tilde\lambda_n-\lambda_n}
{x-2M_n[\tilde Y_x-\tilde Y^2]}
\e
for the $P_{II}$ hierarchy, along with the shift in paramaters (\ref{p2s}).
We note that this derivation, which does not make use of the Schwarzian
derivative but instead relies on the invariance of the quantity $Y_x-Y^2$
under the mapping (\ref{Ytr}), (\ref{Ytr1}), is much simpler, and is much 
more widely applicable, than that presented in \cite{CJP99}.

Special integrals of the $P_{II}$ hierarchy are obtained by setting
coefficients of different powers of $\varphi$ in (\ref{wombat}) to zero 
independently; since $\tilde Y_x-\tilde Y^2=Y_x-Y^2$ we see that this gives
\b
M_n[Y_x-Y^2]-\frac{1}{2}x=0,
\e
which defines solutions of (\ref{p2h}) for $\lambda_n=1/2$. We refer to
\cite{CJP99} for further information on special integrals of the $P_{II}$
hierarchy, and the iteration of $P_{II}$ hierarchy BTs.

\section{B\"acklund transformations for fourth Painlev\'e hierarchies}

We now apply the above approach to our $P_{IV}$ hierarchy (\ref{mod1a});
since
\b
{\bf H}[\bm{\phi}]=\left(
\begin{array}{c} \phi+2p \\ \phi p +p^2+p_x \end{array}
\right),
\e
this reads
\b
\left(\begin{array}{cc} 1 & p \\ 2 & \phi+2p-\partial_x \end{array}\right)
{\bf K}_n\left[\left(
\begin{array}{c} \phi+2p \\ \phi p +p^2+p_x \end{array}
\right)\right]
+\left(\begin{array}{c} e_n \\ f_n \end{array}\right)=
\left(\begin{array}{c} 0 \\ 0 \end{array}\right).
\label{p41}
\e

We now seek, analogously to the case of the $P_{II}$ hierarchy above,
a mapping (BT) between two solutions $\phi$, $p$ and $\tilde\phi$, $\tilde p$
of our $P_{IV}$ hierarchy, of the form
\ba
\phi & = & 2\frac{\varphi_x}{\varphi}+\tilde \phi, \label{tr1a} \\
p & = & -\frac{\varphi_x}{\varphi}+\tilde p, \label{tr1b}
\ea
where
\b
\tilde \phi = -\frac{\varphi_{xx}}{\varphi_x}. \label{tr2}
\e
It then follows that
\b
\phi+2p=\tilde\phi+2\tilde p
\e
and
\b
\phi p +p^2+p_x=\tilde\phi \tilde p +\tilde p^2+\tilde p_x
-\frac{\varphi_{xx}}{\varphi}-\frac{\varphi_x}{\varphi}\tilde\phi
=\tilde\phi \tilde p +\tilde p^2+\tilde p_x,
\e
where the last equality follows from (\ref{tr2}). Thus we see that
the quantities $\phi+2p$ and $\phi p +p^2+p_x$ are invariant under the
mapping (\ref{tr1a}),  (\ref{tr1b}), (\ref{tr2}). Substitution of
(\ref{tr1a}),  (\ref{tr1b}) into (\ref{p41}) therefore gives
\b
\left(\begin{array}{cc} 1 & \tilde p-\frac{\varphi_x}{\varphi} \\ 
2 & \tilde \phi+2\tilde p-\partial_x \end{array}\right)
{\bf K}_n\left[\left(
\begin{array}{c} \tilde \phi+2\tilde p \\ 
\tilde\phi \tilde p +\tilde p^2+\tilde p_x \end{array}
\right)\right]
+\left(\begin{array}{c} e_n \\ f_n \end{array}\right)=
\left(\begin{array}{c} 0 \\ 0 \end{array}\right).
\label{p41s}
\e
Since we assume that $\tilde \phi$, $\tilde p$ are solutions of a second
copy of our $P_{IV}$ hierarchy, but with parameters $\tilde e_n$, 
$\tilde f_n$, i.e.
\b
\left(\begin{array}{cc} 1 & \tilde p \\ 
2 & \tilde \phi+2\tilde p-\partial_x \end{array}\right)
{\bf K}_n\left[\left(
\begin{array}{c} \tilde \phi+2\tilde p \\ 
\tilde \phi \tilde p +\tilde p^2+\tilde p_x \end{array}
\right)\right]
+\left(\begin{array}{c} \tilde e_n \\ \tilde f_n \end{array}\right)=
\left(\begin{array}{c} 0 \\ 0 \end{array}\right),
\label{p41t}
\e
we obtain, by elimination between (\ref{p41s}) and this last,
\ba
\frac{\varphi_x}{\varphi} & = & \frac{e_n-\tilde e_n}{L_n\left[
\left(
\begin{array}{c} \tilde \phi+2\tilde p \\ 
\tilde \phi \tilde p +\tilde p^2+\tilde p_x \end{array}
\right)
\right]}, \label{c1a} \\
\tilde f_n & = & f_n. \label{c1b}
\ea
Equation (\ref{c1a}) must be compatible with (\ref{tr2}), or equivalently
with the Riccati equation
\b
\left(\frac{\varphi_x}{\varphi}\right)_x+
\left(\frac{\varphi_x}{\varphi}\right)^2
+\tilde \phi \left(\frac{\varphi_x}{\varphi}\right)=0;
\label{ric1}
\e
substituting (\ref{c1a}) into (\ref{ric1}) gives
\b
(\tilde\phi-\partial_x)L_n\left[
\left(
\begin{array}{c} \tilde \phi+2\tilde p \\ 
\tilde \phi \tilde p +\tilde p^2+\tilde p_x \end{array}
\right)
\right]
+(e_n-\tilde e_n)=0,
\e
and comparing this last with (\ref{p41t}) we see that we must have
$\tilde e_n = \tilde f_n-e_n$ and so 
\b
\tilde e_n = f_n-e_n.
\label{c1c}
\e

Thus we have for our $P_{IV}$ hierarchy (\ref{p41}) the BT
\ba
\phi & = & \tilde\phi+2\frac{e_n-\tilde e_n}{L_n\left[
\left(
\begin{array}{c} \tilde \phi+2\tilde p \\ 
\tilde \phi \tilde p +\tilde p^2+\tilde p_x \end{array}
\right)
\right]}, \label{bt4a} \\
p & = & \tilde p-\frac{e_n-\tilde e_n}{L_n\left[
\left(
\begin{array}{c} \tilde \phi+2\tilde p \\ 
\tilde \phi \tilde p +\tilde p^2+\tilde p_x \end{array}
\right)
\right]},  \label{bt4b}
\ea
along with the shifts in parameters given by (\ref{c1b}) and (\ref{c1c}).

We now consider deriving BTs for the $P_{IV}$ hierarchy (\ref{mod1b}); since
\b
{\bf I}[\bm{\psi}]=\left(
\begin{array}{c} \psi-2s \\ -\psi s +s^2-s_x \end{array}
\right),
\e
this reads
\b
\left(\begin{array}{cc} 1 & -s \\ -2 & -\psi+2s+\partial_x \end{array}\right)
{\bf K}_n\left[\left(
\begin{array}{c} \psi-2s \\ -\psi s +s^2-s_x \end{array}
\right)\right]
+\left(\begin{array}{c} l_n \\ m_n \end{array}\right)=
\left(\begin{array}{c} 0 \\ 0 \end{array}\right).
\label{p42}
\e
Seeking a BT in the form
\ba
\psi & = & 2\frac{\varphi_x}{\varphi}+\hat \psi, \label{tr2a} \\
s & = & \frac{\varphi_x}{\varphi}+\hat s, \label{tr2b}
\ea
where
\b
\hat \psi = -\frac{\varphi_{xx}}{\varphi_x}, \label{trc2}
\e
and where $\hat\psi$, $\hat s$ are solutions of our $P_{IV}$ hierarchy 
for parameter values $\hat l_n$, $\hat m_n$,
\b
\left(\begin{array}{cc} 1 & -\hat s 
\\ -2 & -\hat \psi+2\hat s+\partial_x \end{array}\right)
{\bf K}_n\left[\left(
\begin{array}{c} \hat \psi-2\hat s 
\\ -\hat \psi \hat s +\hat s^2-\hat s_x \end{array}
\right)\right]
+\left(\begin{array}{c} \hat l_n \\ \hat m_n \end{array}\right)=
\left(\begin{array}{c} 0 \\ 0 \end{array}\right),
\label{p42t}
\e
then yields
\ba
\psi & = & \hat\psi+2\frac{l_n-\hat l_n}{L_n\left[
\left(
\begin{array}{c} \hat \psi-2\hat s \\ 
-\hat \psi \hat s +\hat s^2-\hat s_x \end{array}
\right)
\right]}, \label{bt2a} \\
s & = & \hat s+\frac{l_n-\hat l_n}{L_n\left[
\left(
\begin{array}{c} \hat \psi-2\hat s \\ 
-\hat \psi \hat s +\hat s^2-\hat s_x \end{array}
\right)
\right]}, \label{bt2b}
\ea
for the shift in parameter values
\ba
\hat m_n & = & m_n, \label{bt2c} \\
\hat l_n & = & -m_n -l_n. \label{bt2d}
\ea
We note that the BT (\ref{bt2a})---(\ref{bt2d}) follows immediately
from (\ref{bt4a}), (\ref{bt4b}), (\ref{c1b}), (\ref{c1c}) under
$(\phi,p,e_n,f_n)\rightarrow (\psi,-s,l_n,-m_n)$, which maps the $P_{IV}$
hierarchy (\ref{p41}) into the $P_{IV}$ hierarchy (\ref{p42}). However,
this mapping does not leave the $P_{IV}$ equation in standard form, since
(\ref{p4p}) is mapped to (\ref{p4s}), and these two equations we identify 
with $P_{IV}$ by setting $g_1=2$ and $g_1=-2$ respectively. We return to
this point later.

We now briefly consider special integrals. We see that setting coefficients
of different powers of $\varphi$ in (\ref{p41s}) to zero independently gives,
using the fact that $\tilde \phi+2\tilde p=\phi+2p$ and $\tilde \phi\tilde p
+\tilde p^2+\tilde p_x=\phi p +p^2+p_x$,
\b
L_n\left[
\left(
\begin{array}{c} \phi+2p \\ \phi p +p^2+p_x \end{array}
\right)
\right]=0,
\label{si1a}
\e
which then defines solutions of (\ref{p41}) provided that
\b
K_n\left[
\left(
\begin{array}{c} \phi+2p \\ \phi p +p^2+p_x \end{array}
\right)
\right]+e_n=0
\label{si1b}
\e
and
\b
f_n=2e_n.
\label{sip1}
\e
In the same way, at the same point in the derivation of the BT
(\ref{bt2a})---(\ref{bt2d}), setting coefficients of different 
powers of $\varphi$ to zero independently, and using the fact
that $\hat\psi-2\hat s = \psi-2s$ and $-\hat \psi\hat s 
+\hat s^2 -\hat s_x =-\psi s +s^2 -s_x$, gives
\b
L_n\left[
\left(
\begin{array}{c} \psi-2s \\ 
-\psi s +s^2-s_x \end{array}
\right)
\right]=0,
\label{si2a}
\e
which then defines solutions of (\ref{p42}) provided that
\b
K_n\left[
\left(
\begin{array}{c} \psi-2s \\ 
-\psi s +s^2-s_x \end{array}
\right)
\right]+l_n=0
\label{si2b}
\e
and
\b
m_n=-2l_n.
\label{sip2}
\e
Again, just as for our BTs, we have the mapping 
$(\phi,p,e_n,f_n)\rightarrow (\psi,-s,l_n,-m_n)$
between these special integrals. See however the
discussion in the next Section.

\section{Identification of B\"acklund transformations}

We now turn to the identification of the BTs obtained in the previous
section. We will give to BTs for our $P_{IV}$ hierarchies the same names
as are given to the case $n=1$, i.e.\ to the $P_{IV}$ equation itself.
First of all we fix the identification of parameters in our hierachies
with the parameters $\alpha$ and $\beta$ in $P_{IV}$ when written as
\b
Q_{xx}=\frac{1}{2}\frac{Q_x^2}{Q}+\frac{3}{2}Q^3+4xQ^2+2\left(x^2-\alpha
\right)Q-\frac{1}{2}\frac{\beta^2}{Q}.
\label{p4Q}
\e
We take this last as the standard form of $P_{IV}$ in order to simplify the 
writing of parameter shifts for its BTs. We note that, since 
$\beta\rightarrow-\beta$
is a discrete symmetry of (\ref{p4Q}), in BTs of $P_{IV}$ we can always
replace parameters corresponding to $\beta$ by $\pm\beta$.
 
We begin with the hierarchy (\ref{mod1a}), or (\ref{p41}),
\b
\left({\bf H}'[\bm{\phi}]\right)^{\dagger}{\bf K}_n[{\bf H}[\bm{\phi}]]
+(e_n,f_n)^T=0.
\label{p4I}
\e
In the case $n=1$ this gives the system (\ref{m21a}), (\ref{m21b}) and,
after eliminating $\phi$, equation (\ref{p4p}). In order to identify
this last with equation (\ref{p4Q}) we now set, for the entire hierarchy 
(\ref{p4I}),
\ba
g_n & = & 2, \label{id1a} \\
e_n & = & B_n, \label{id1b} \\
f_n & = & \frac{1}{2}(2A_n+3B_n+2),  \label{id1c}
\ea
and similarly for the parameters $\tilde e_n$ and $\tilde f_n$ in the hierarchy
(\ref{p41t}).

We then have for the hierarchy (\ref{p4I}) the BT (\ref{bt4a}), (\ref{bt4b}),
with the corresponding shift on parameters (\ref{c1b}), (\ref{c1c}), i.e.
\ba
\tilde A_n & = & -\frac{1}{4}(2A_n-3B_n+6), \label{par1a} \\
\tilde B_n & = & \frac{1}{2}(2A_n+B_n+2). \label{par1b}
\ea
In the case $n=1$ this BT reads
\ba
\phi & = & \tilde \phi +\frac{B_1-2A_1-2}{\tilde\phi+2\tilde p +2x}, \\
p & = & \tilde p -\frac{1}{2}\frac{B_1-2A_1-2}{\tilde\phi+2\tilde p +2x}.
\ea
Eliminating $\tilde\phi$ we obtain a BT for $P_{IV}$ (\ref{p4p}) itself,
\b
p=\tilde p +\frac{(B_1-2A_1-2)\tilde p}
{\tilde p_x-\tilde p^2-2x\tilde p+B_1/2+A_1+1}.
\label{pBT}
\e
This BT for $P_{IV}$, along with the parameter shift (\ref{par1a}),
(\ref{par1b}) (for $n=1$), is often referred to as the ``double dagger''
($t^\ddagger$) BT (see \cite{BCH95,GJP99}). For this reason we refer to the 
BT (\ref{bt4a}), (\ref{bt4b}), together with the parameter shifts 
(\ref{par1a}), (\ref{par1b}), as the $t^\ddagger$ BT for the $P_{IV}$ 
hierarchy (\ref{p4I}).

We now turn to the hierarchy (\ref{mod1b}), or (\ref{p42}),
\b
\left({\bf I}'[\bm{\psi}]\right)^{\dagger}{\bf K}_n[{\bf I}[\bm{\psi}]]
+(l_n,m_n)^T=0,
\label{p4II}
\e
In the case $n=1$ this gives the system (\ref{n21a}), (\ref{n21b}) and,
after eliminating $\psi$, equation (\ref{p4s}). In order to identify
equation (\ref{p4s}) with (\ref{p4Q}) we now set, for the entire 
hierarchy (\ref{p4II}),
\ba
g_n & = & -2, \label{id2a} \\
l_n & = & b_n, \label{id2b} \\
m_n & = & -\frac{1}{2}(2a_n+3b_n-2), \label{id2c}
\ea
and analogously for the parameters $\hat l_n$ and $\hat m_n$ in the 
hierarchy (\ref{p42t}).

We have for the hierarchy (\ref{p4II}) the BT (\ref{bt2a}), (\ref{bt2b}),
with the corresponding shift on parameters (\ref{bt2c}), (\ref{bt2d}), i.e.
\ba
\hat a_n & = & -\frac{1}{4}(2a_n-3b_n-6), \label{par2a} \\
\hat b_n & = & \frac{1}{2}(2a_n+b_n-2). \label{par2b}
\ea
In the case $n=1$ this BT reads
\ba
\psi & = & \hat \psi +\frac{b_1-2a_1+2}{\hat\psi-2\hat s -2x}, \\
s & = & \hat s +\frac{1}{2}\frac{b_1-2a_1+2}{\hat\psi-2\hat s -2x},
\ea
and eliminating $\hat\psi$ we obtain a BT for $P_{IV}$ (\ref{p4s}) itself,
\b
s=\hat s +\frac{(2a_1-b_1-2)\hat s}
{\hat s_x+\hat s^2+2x\hat s-b_1/2-a_1+1}.
\label{sBT}
\e
This BT for $P_{IV}$, along with the parameter shift (\ref{par2a}),
(\ref{par2b}) (for $n=1$), is often refereed to as the ``dagger''
($\tau^\dagger$) BT (see \cite{BCH95,GJP99}); it is for this reason that
we refer to the BT (\ref{bt2a}), (\ref{bt2b}), together with the parameter 
shifts (\ref{par2a}), (\ref{par2b}), as the $\tau^\dagger$ BT for the $P_{IV}$ 
hierarchy (\ref{p4II}). Here we use the letter ``$t$'' (e.g.\ $t^\ddagger$) for 
BTs related to the hierarchy (\ref{p4I}), and ``$\tau$'' (e.g.\ $\tau^\dagger$) 
for BTs related to the hierarchy (\ref{p4II}).

Finally we recall that we also have, as detailed in Section Four, special 
integrals for our $P_{IV}$ hierarchies. Thus we have the special integral 
system (\ref{si1a})---(\ref{si1b}), with parameters satisfying (\ref{sip1}),
for the hierarchy (\ref{p4I}), where we now impose the identification
(\ref{id1a})---(\ref{id1c}). Similarly we have the special integral system 
(\ref{si2a})---(\ref{si2b}), with parameters satisfying (\ref{sip2}), for 
the hierarchy (\ref{p4II}), now imposing (\ref{id2a})---(\ref{id2c}).

In the case $n=1$, with the identification (\ref{id1a})---(\ref{id1c}), our 
special integral system (\ref{si1a})---(\ref{si1b}) for the hierarchy 
(\ref{p4I}) reads
\ba
\phi+2p+2x & = & 0, \\
p_x+\phi p +p^2+B_1 & = & 0,
\ea
for parameters satisfying (\ref{sip1}), i.e.
\b
B_1=2A_1+2.
\label{sip1a}
\e
Thus we obtain the special integral of $P_{IV}$ (\ref{p4p}),
\b
p_x-p^2-2xp+B_1=0,
\e
where the parameters $A_1$ and $B_1$ of $P_{IV}$ satisfy (\ref{sip1a}).

On the other hand, the special integral system (\ref{si2a})---(\ref{si2b}) 
for the hierarchy  (\ref{p4II}), with the identification 
(\ref{id2a})---(\ref{id2c}), reads for $n=1$,
\ba
\psi-2s-2x & = & 0, \\
s_x+\psi s -s^2-b_1 & = & 0,
\ea
with parameters satisfying (\ref{sip2}), i.e.
\b
b_1=2a_1-2.
\label{sip2a}
\e
Eliminating $\psi$ then gives the special integral of $P_{IV}$ (\ref{p4s}),
\b
s_x+s^2+2xs-b_1=0,
\e
for parameters $a_1$ and $b_1$ of $P_{IV}$ satisfying (\ref{sip2a}).

We note that the identifications of parameters (\ref{id1a})---(\ref{id1c})
and (\ref{id2a})---(\ref{id2c}) mean that we no longer have the simple
mapping $(\phi,p,e_n,f_n)\rightarrow (\psi,-s,l_n,-m_n)$ between the 
hierarchies (\ref{p4I}) and (\ref{p4II}); consider for example the systems 
obtained for $n=1$, (\ref{m21a}), (\ref{m21b}) and (\ref{n21a}), (\ref{n21b}). 
Thus the BTs and special integrals obtained here are no longer equivalent 
under this mapping. The question of whether a mapping can be found under which
they are equivalent is discussed in Section Seven.

\section{Further B\"acklund transformations for our $P_{IV}$ hierarchies}

Thus far we have found the BTs $t^\ddagger$ and $\tau^\dagger$ for our $P_{IV}$
hierarchies. However, as is well known, for $P_{IV}$ itself, these BTs can 
be written as compositions of other BTs, referred to in the literature 
\cite{BCH95,GJP99} as the ``tilde'' $\left(\tilde t/\tilde \tau\right)$ and 
``hat'' 
$\left(\hat t/\hat \tau\right)$ BTs. We now show how this pattern of BTs for $P_{IV}$
can be extended to our $P_{IV}$ hierarchies.

It turns out, quite remarkably, that this can be done by considering the
Miura maps between (\ref{p4I}) and (\ref{mod1}), and (\ref{p4II}) and
(\ref{mod1}), as given in (\ref{seq}). Let us begin with the Miura
transformation between (\ref{p4I}) and (\ref{mod1}), as given by (\ref{mm2a}).
We recall that for $n=1$ (\ref{mod1}) yields equation (\ref{p4y}). We take
the lower sign in (\ref{p4y}) and now fix the relationship between our 
parameters $c_n$, $d_n$ and $\alpha_n$, $\beta_n$, for the entire hierarchy
(\ref{mod1}), as
\ba
g_n & = & 2, \label{id3a} \\
c_n & = & \frac{1}{2}(2-2\alpha_n-\beta_n), \\
d_n & = & \beta_n, \label{id3c}
\ea
and similarly for a second copy of our hierarchy (\ref{mod1}) in $\tilde U$,
$\tilde V$ with parameters $\tilde c_n$, $\tilde d_n$, or equivalently
$\tilde\alpha_n$, $\tilde\beta_n$. 

Since we have
\ba
c_n & = & e_n, \\
d_n & = & f_n-2e_n
\ea
and similarly for parameters $\tilde c_n$, $\tilde d_n$, $\tilde e_n$, and
$\tilde f_n$, we obtain the following BTs and parameter shifts:
\ba
\tilde U & = & \tilde \phi+2\tilde p, \label{til1} \\
\tilde V & = & \tilde p, \label{til2}
\ea
with
\ba
\tilde A_n & = & -\frac{1}{4}(2+2\tilde \alpha_n-3\tilde \beta_n), 
\label{til3} \\
\tilde B_n & = & \frac{1}{2}(2-2\tilde \alpha_n-\tilde \beta_n);
\label{til4} 
\ea
and
\ba
\phi & = & U-2V, \label{hat1} \\
p & = & V,  \label{hat2}
\ea
with
\ba
\alpha_n & = & \frac{1}{4}(2-2A_n-3B_n),  \label{hat3} \\
\beta_n & = & \frac{1}{2}(2+2A_n-B_n). \label{hat4}
\ea

For the case $n=1$, the first of these, when written as a BT between 
two copies of $P_{IV}$ --- (\ref{p4p}) in $\tilde p$, $\tilde A_1$,
$\tilde B_1$, and (\ref{p4y}) in $\tilde y$, $\tilde\alpha_1$, 
$\tilde\beta_1$, where $\tilde y=-\tilde U-2x$ ---  reads
\b
\tilde y=\frac{\tilde p_x-\tilde p^2-2x\tilde p
              +1-\tilde\alpha_1-\tilde\beta_1/2}{2\tilde p},
\label{tilp}
\e
which, together with the parameter shifts (\ref{til3}), (\ref{til4})
defines precisely the BT $\tilde t$.

The second of the above BTs, in the case $n=1$, when written as a BT
between (\ref{p4p}) in $p$, $A_1$, $B_1$ and (\ref{p4y}) in $y$,
$\alpha_1$ and $\beta_1$, where $y=-U-2x$, reads
\b
p=-\frac{y_x+y^2+2xy+1+A_1-B_1/2}{2y}
\label{hatp}
\e
which, together with the parameter shifts (\ref{hat3}), (\ref{hat4})
defines precisely the BT $\hat t$.

We thus define the BTs (\ref{til1}), (\ref{til2}) and (\ref{hat1}), 
(\ref{hat2}), with parameter shifts (\ref{til3}), (\ref{til4}) and
(\ref{hat3}), (\ref{hat4}) respectively, as $\tilde t$ and $\hat t$
BTs for our $P_{IV}$ hierarchies.

We now define the additional BT $S$ by 
$S=(\hat t)^{-1}\circ t^{\ddagger}\circ (\tilde t)^{-1}$. A simple
calculation gives this BT $S$ as
\ba
U & = & \tilde U, \label{sig1} \\
V & = & \tilde V +\frac{d_n}{L_n\left[\left(\begin{array}{c} \tilde U \\
\tilde U\tilde V-\tilde V^2+\tilde V_x \end{array}\right)\right]},
\label{sig2}
\ea
with the change of parameters
\ba
\tilde c_n & = & c_n+d_n, \\
\tilde d_n & = & -d_n,
\ea
or equivalently
\ba
\tilde \alpha_n & = & \alpha_n, \\
\tilde \beta_n & = & -\beta_n. \label{sig6}
\ea
For the case $n=1$, for equation (\ref{p4y}), this BT $S$ reads
\b
y = \tilde y, \qquad{} \tilde \alpha_1 = \alpha_1,  \qquad{}
\tilde \beta_1 = -\beta_1,
\e
and we recover the well known relation 
$t^{\ddagger}=\hat t\circ S\circ\tilde t$ for $P_{IV}$ BTs. Our BT
$S$ (\ref{sig1})---(\ref{sig6}) then allows us to extend this 
decomposition of the $BT$ $t^{\ddagger}$ as 
$t^{\ddagger}=\hat t\circ S\circ\tilde t$ from $P_{IV}$ itself to 
our $P_{IV}$ hierarchies. This pattern of BTs, obtained here using the
Miura map ${\bf U}={\bf \Phi}[\bm{\phi}]$ of the DWW hierarchy (which 
defines the BTs $\hat t$ and $\tilde t$), can be seen in Figure One. It
is interesting that this Miura map, a simple linear map
when considered as a mapping between $\bm{\phi}$ and ${\bf U}$, gives rise
to BTs of our hierarchies: however, as we have seen above for $n=1$ ($P_{IV}$),
when considered as a mapping between components of our hierarchies,
it is no longer a linear map.

We recall that for $n=1$ the second component $V$ of the system (\ref{mod1})
also defines a copy of $P_{IV}$ (\ref{p4w}). Using the identification
(\ref{id1a})---(\ref{id1c}), where as usual $c_1=e_1$ and $d_1=f_1-2e_1$,
we obtain that equation (\ref{p4w}), with the upper choice of sign, is
a copy of equation (\ref{p4p}), with $w=p$ and the same parameters $A_1$ and
$B_1$. Thus, in Figure One, when tracing for $n=1$ the action of our BTs
over individual components, we see that the auto-BT for (\ref{p4p}) must be
the same as the auto-BT for equation (\ref{p4w}). This last is as given by
(\ref{sig2}), and reads (with $V=w$, $\tilde V=\tilde w$, and eliminating
$\tilde U$),
\b
w=\tilde w +\frac{(B_1-2A_1-2)\tilde w}
{\tilde w_x-\tilde w^2-2x\tilde w+B_1/2+A_1+1},
\e
with parameter shifts
\ba
\tilde A_1=-\frac{1}{4}(2A_1-3B_1+6), \\
\tilde B_1=\frac{1}{2}(2A_1+B_1+2).
\ea
Thus we see that this BT is exactly the same as that for (\ref{p4p}),
i.e.\ (\ref{pBT}) and (\ref{par1a}), (\ref{par1b}), with $n=1$. That is, 
for $n=1$, (\ref{sig2}) is the $t^\ddagger$ BT (from $\tilde w$ to $w$). 

From the above it also follows that the equation obtained when
eliminating $\tilde U$ from the system (\ref{UV1}), (\ref{UV2})
written in terms of $\tilde U$, $\tilde V$, $\tilde c_1$, $\tilde d_1$,
i.e.
\b
\tilde U=-\frac{\tilde V_x-\tilde V^2+2x\tilde V+\tilde c_1}{2\tilde V},
\label{rel1}
\e
corresponds to the $\tilde t$ BT from (\ref{p4w}), with upper sign, to 
(\ref{p4y}), with lower sign. In the same way, the equation obtained
when eliminating $V$ from the system (\ref{UV1}), (\ref{UV2}),
\b
V=-\frac{U_x-U^2-2xU+2-d_1}{2U+4x},
\label{rel2}
\e
corresponds to the $\hat t$ BT from (\ref{p4y}), with lower sign, to
(\ref{p4w}), with upper sign. This then gives one identification of
equations (\ref{rel1}) and (\ref{rel2}) (another is made later).

We now turn to our $\tau^\dagger$ BT. We recall once again the Miura maps
(\ref{seq}) and in particular the Miura transformation between (\ref{p4II}) 
and (\ref{mod1}), as given by (\ref{mm2b}). For $n=1$ (\ref{mod1}) yields 
equation (\ref{p4y}); we take the upper sign in (\ref{p4y}) and change
the relationship between our parameters (now labelled $\bar c_n$, $\bar d_n$ 
and $\bar \alpha_n$, $\bar \beta_n$, corresponding to variables $\bar U$, 
$\bar V$), to the following, again for the entire hierarchy (\ref{mod1}):
\ba
g_n & = & -2, \label{id4a} \\
\bar c_n & = & -\frac{1}{2}(2+2\bar \alpha_n+\bar \beta_n), \\
\bar d_n & = & \bar \beta_n, \label{id4c}
\ea
and similarly for a second copy of our hierarchy (\ref{mod1}) in $\hat U$,
$\hat V$ with parameters $\hat c_n$, $\hat d_n$, or equivalently
$\hat\alpha_n$, $\hat\beta_n$. 

Since we have
\ba
\bar c_n & = & l_n, \\
\bar d_n & = & -m_n-2l_n
\ea
and similarly for parameters $\hat c_n$, $\hat d_n$, $\hat l_n$, and
$\hat m_n$, we obtain the following BTs and parameter shifts:
\ba
\hat U & = & \hat \psi-2\hat s, \label{til1a} \\
\hat V & = & -\hat s, \label{til2a}
\ea
with
\ba
\hat a_n & = & \frac{1}{4}(2-2\hat \alpha_n+3\hat \beta_n), 
\label{til3a} \\
\hat b_n & = & -\frac{1}{2}(2+2\hat \alpha_n+\hat \beta_n);
\label{til4a} 
\ea
and
\ba
\psi & = & \bar U-2\bar V, \label{hat1a} \\
s & = & -\bar V,  \label{hat2a}
\ea
with
\ba
\bar\alpha_n & = & -\frac{1}{4}(2+2a_n+3b_n),  \label{hat3a} \\
\bar\beta_n & = & -\frac{1}{2}(2-2a_n+b_n). \label{hat4a}
\ea

For the case $n=1$, the first of these, when written as a BT between 
two copies of $P_{IV}$ --- (\ref{p4s}) in $\hat s$, $\hat a_1$,
$\hat b_1$, and (\ref{p4y}) in $\hat y$, $\hat\alpha_1$, 
$\hat\beta_1$, where $\hat y=\hat U-2x$ ---  reads
\b
\hat y=-\frac{\hat s_x+\hat s^2+2x\hat s
              +1+\hat\alpha_1+\hat\beta_1/2}{2\hat s},
\e
which, together with the parameter shifts (\ref{til3a}), (\ref{til4a})
defines the BT $\hat\tau$ (we recall, when comparing to the BT 
(\ref{hatp}) with parameter shifts, (\ref{hat3}), (\ref{hat4}) for
$n=1$ --- also identified as a ``hat'' BT, $\hat t$ ---  the invariance 
of $P_{IV}$ (\ref{p4Q}) under $\beta\rightarrow-\beta$).

The second of the above BTs, in the case $n=1$, when written as a BT
between (\ref{p4s}) in $s$, $a_1$, $b_1$ and (\ref{p4y}) in $\bar y$,
$\bar\alpha_1$ and $\bar\beta_1$, where $\bar y=\bar U-2x$, reads
\b
s=\frac{\bar y_x-\bar y^2-2x\bar y+1-a_1+b_1/2}{2\bar y}
\label{tiltau}
\e
which, together with the parameter shifts (\ref{hat3a}), (\ref{hat4a})
defines the BT $\tilde\tau$ (again we recall, when comparing to  
(\ref{tilp}) with parameter shifts, (\ref{til3}), (\ref{til4}) for
$n=1$ --- also identified as a ``tilde'' BT, $\tilde t$ ---  the invariance 
of $P_{IV}$ (\ref{p4Q}) under $\beta\rightarrow-\beta$).

We thus define the BTs (\ref{til1a}), (\ref{til2a}) and (\ref{hat1a}), 
(\ref{hat2a}), with parameter shifts (\ref{til3a}), (\ref{til4a}) and
(\ref{hat3a}), (\ref{hat4a}) respectively, as $\hat\tau$ and $\tilde\tau$
BTs for our $P_{IV}$ hierarchies.

We now define the additional BT $\sigma$ by 
$\sigma=(\tilde\tau)^{-1}\circ \tau^{\dagger}\circ (\hat\tau)^{-1}$. A simple
calculation gives this BT $\sigma$ as
\ba
\bar U & = & \hat U, \label{sigs1} \\
\bar V & = & \hat V +\frac{\bar d_n}{L_n\left[\left(\begin{array}{c} \hat U \\
\hat U\hat V-\hat V^2+\hat V_x \end{array}\right)\right]},
\label{sigs2}
\ea
with the change of parameters
\ba
\hat c_n & = & \bar c_n+\bar d_n, \\
\hat d_n & = & -\bar d_n,
\ea
or equivalently
\ba
\hat \alpha_n & = & \bar\alpha_n, \\
\hat \beta_n & = & -\bar\beta_n. \label{sigs6}
\ea
For the case $n=1$, for equation (\ref{p4y}), this BT $\sigma$ reads
\b
\bar y = \hat y, \qquad{} \hat \alpha_1 = \bar\alpha_1,  \qquad{}
\hat \beta_1 = -\bar\beta_1,
\e
and we recover the well known relation 
$\tau^{\dagger}=\tilde \tau\circ \sigma\circ\hat \tau$ for $P_{IV}$ BTs. Our 
BT $\sigma$ (\ref{sigs1})---(\ref{sigs6}) then allows us to extend this 
decomposition of the $BT$ $\tau^{\dagger}$ as 
$\tau^{\dagger}=\tilde \tau\circ \sigma\circ\hat \tau$ from $P_{IV}$ itself to 
our $P_{IV}$ hierarchies. This pattern of BTs, obtained here using the Miura
map ${\bf U}={\bf \Psi}[\bm{\psi}]$ of the DWW hierarchy, can be seen in Figure 
Two. Again we note that what is a simple linear map, when considered as a 
mapping between $\bm{\psi}$ and ${\bf U}$, gives rise to BTs of our hierarchies
(the BTs $\hat\tau$ and $\tilde\tau$), but that, once again, and as we have 
seen for $n=1$ ($P_{IV}$), when considered as a mappping between components of 
our hierarchies, it is no longer such a trivial mapping.

We recall that for $n=1$ the second component $\bar V$ of the system 
(\ref{mod1}), now written in terms of $\bar U$, $\bar V$ and with
coefficients $\bar c_1$, $\bar d_1$, also defines a copy of $P_{IV}$ via
$\bar V=-\bar w$. This copy of $P_{IV}$ is (\ref{p4w}) in $\bar w$, 
$\bar c_1$ and $\bar d_1$, and with lower choice of sign. We thus obtain 
that this copy of (\ref{p4w}), where as usual $\bar c_1=l_1$ and 
$\bar d_1=-m_1-2l_1$, and using the identification (\ref{id2a})---(\ref{id2c}), 
is a copy of equation (\ref{p4s}), with $w=s$ and the same parameters $a_1$ and 
$b_1$. Thus, in Figure Two, when tracing for $n=1$ the action of our BTs over 
individual components, we see that the auto-BT for (\ref{p4s}) must be the same 
as the auto-BT for this copy of equation (\ref{p4w}). This last is as given by 
(\ref{sigs2}), and reads (with $\bar V=-\bar w$, $\hat V=-\hat w$, and 
eliminating $\hat U$),
\b
\bar w=\hat w +\frac{(2a_1-b_1-2)\hat w}
{\hat w_x+\hat w^2+2x\hat w-b_1/2-a_1+1},
\e
with parameter shifts
\ba
\hat a_1=-\frac{1}{4}(2a_1-3b_1-6), \\
\hat b_1=\frac{1}{2}(2a_1+b_1-2).
\ea
Thus we see that this BT is exactly the same as that for (\ref{p4s}),
i.e.\ (\ref{sBT}) and (\ref{par2a}), (\ref{par2b}), with $n=1$. That is, 
for $n=1$, (\ref{sigs2}) is the $\tau^\dagger$ BT (from $\hat w$ to $\bar w$). 

From the above it also follows that the equation obtained when
eliminating $\hat U$ from the system (\ref{UV1}), (\ref{UV2}),
written in terms of $\hat U$, $\hat V$, $\hat c_1$, $\hat d_1$,
i.e.
\b
\hat U=-\frac{\hat V_x-\hat V^2+2x\hat V+\hat c_1}{2\hat V},
\label{rel3}
\e
corresponds to the $\hat \tau$ BT from (\ref{p4w}), with lower 
sign, to (\ref{p4y}), with upper sign. In the same way, the 
equation obtained when eliminating $\bar V$ in the system 
(\ref{UV1}), (\ref{UV2}), written in terms of $\bar U$, $\bar V$, 
$\bar c_1$, $\bar d_1$, i.e.
\b
\bar V=-\frac{\bar U_x-\bar U^2-2x\bar U+2-\bar d_1}{2\bar U+4x},
\label{rel4}
\e
corresponds to the $\tilde \tau$ BT from (\ref{p4y}), with upper 
sign, to (\ref{p4w}), with lower sign. Thus we have a different
identification of the equations (\ref{rel1}) and (\ref{rel2}).

We note that the transformation induced on our original variables
$u$ and $v$ by $S$ and $\sigma$ is just the identity together with
$(\alpha_n,\beta_n)\rightarrow(\alpha_n,-\beta_n)$ (a discrete 
symmetry of the hierarchy (\ref{si1}), (\ref{si2})).

\section{A mapping between our sequences of $P_{IV}$ hierarchies}

In this section we consider a mapping between our two different sequences
of fourth Painlev\'e hierarchies. These two different sequences are defined
by the choice $g_n=2$ or $g_n=-2$. As we noted earlier, this then means that
we no longer have the mapping $(\phi,p,e_n,f_n)\rightarrow(\psi,-s,l_n,-m_n)$
between the hierarchies (\ref{mod1a}) and (\ref{mod1b}). However, we still
have another mapping between these two hierarchies.

Consider the hierarchy (\ref{mod1a}), with the identifications 
(\ref{id1a})---(\ref{id1c}), i.e.
\b
\left({\bf H}'[\bm{\phi}]\right)^{\dagger}{\bf K}_n[{\bf H}[\bm{\phi}]]
+(e_n,f_n)^T=0,
\label{betr}
\e
where ($g_n=2$)
\b
{\bf K}_n[{\bf H}[\bm{\phi}]]={\bf L}_n[{\bf H}[\bm{\phi}]]
+\sum_{i=1}^{n-1}h_i{\bf L}_i[{\bf H}[\bm{\phi}]]
+2\left(\begin{array}{c} 0 \\ x \end{array}\right).
\e
We now consider the scaling transformation $(\phi,p,x)=
(\lambda^{-1} \psi,-\lambda^{-1} s,\lambda\xi)$. This then gives
\b
{\bf L}_i[{\bf H}[\bm{\phi}]]=
\left(\begin{array}{cc}
\frac{1}{\lambda^{i+1}} & 0 \\ 0 & \frac{1}{\lambda^{i}}
\end{array}\right)
{\bf L}_i[{\bf I}[\bm{\psi}]],
\label{tra1}
\e
where in the right-hand-side of this last, derivatives are w.r.t.\ $\xi$.
Thus
\ba
{\bf K}_n[{\bf H}[\bm{\phi}]] & = &
\left(\begin{array}{cc}
\frac{1}{\lambda} & 0 \\ 0 & 1
\end{array}\right)\frac{1}{\lambda^{n}}\left[
{\bf L}_n[{\bf I}[\bm{\psi}]]
+\sum_{i=1}^{n-1}h_i\lambda^{n-i}{\bf L}_i[{\bf I}[\bm{\psi}]]
+2\lambda^{n+1}\left(\begin{array}{c} 0 \\ \xi \end{array}\right)
\right] \nonumber \\
 & = & \left(\begin{array}{cc}
\frac{1}{\lambda} & 0 \\ 0 & 1
\end{array}\right)\frac{1}{\lambda^{n}}\left[
{\bf L}_n[{\bf I}[\bm{\psi}]]
+\sum_{i=1}^{n-1}H_i{\bf L}_i[{\bf I}[\bm{\psi}]]
-2\left(\begin{array}{c} 0 \\ \xi \end{array}\right)
\right],
\ea
where we have chosen $\lambda$ such that $\lambda^{n+1}=-1$,
and have set $h_i\lambda^{n-i}=H_i$. Since also
\b
\left({\bf H}'[\bm{\phi}]\right)^{\dagger}=
\left(\begin{array}{cc}
1 & 0 \\ 0 & -1
\end{array}\right)
\left({\bf I}'[\bm{\psi}]\right)^{\dagger}
\left(\begin{array}{cc}
1 & 0 \\ 0 & \frac{1}{\lambda}
\end{array}\right),
\e
our hierarchy (\ref{mod1a}), i.e.\ (\ref{betr}), becomes
\b
\left({\bf I}'[\bm{\psi}]\right)^{\dagger}
\left[
{\bf L}_n[{\bf I}[\bm{\psi}]]
+\sum_{i=1}^{n-1}H_i{\bf L}_i[{\bf I}[\bm{\psi}]]
-2\left(\begin{array}{c} 0 \\ \xi \end{array}\right)
\right]+\left(\begin{array}{c}
-e_n \\ f_n \end{array}\right)=
\left(\begin{array}{c} 0 \\ 0 \end{array}\right),
\e
which, identifying $e_n=-l_n$ and $f_n=m_n$, is precisely the hierarchy 
(\ref{mod1b}), i.e.
\b
\left({\bf I}'[\bm{\psi}]\right)^{\dagger}{\bf K}_n[{\bf I}[\bm{\psi}]]
+(l_n,m_n)^T=0,
\e
with the identification (\ref{id2a})---(\ref{id2c}) ($g_n=-2$). Thus we have 
a BT from the hierarchy (\ref{mod1b}) with $g_n=-2$ to the hierarchy 
(\ref{mod1a}) with $g_n=2$, given by
\ba
(\phi,p,x) & = &
(\lambda^{-1} \psi,-\lambda^{-1} s,\lambda\xi), \\
l_n & = & -e_n, \\
m_n & = & f_n, \\
H_i & = & h_i\lambda^{n-i},
\ea
where $\lambda^{n+1}=-1$.
We refer to this BT as the transformation $T$. We now consider
the composition $T^{-1}\circ t^\ddagger \circ T$, i.e.\ 
$(\hat\psi,\hat s,\hat l_n,\hat m_n)\rightarrow
(\tilde\phi,\tilde p,\tilde e_n,\tilde f_n)\rightarrow
(\phi,p,e_n,f_n)\rightarrow (\psi,s,l_n,m_n)$, which gives an
auto-BT of the hierarchy (\ref{mod1b}). It turns out that this
auto-BT is precisely $\tau^\dagger$. That is, we have the relation
\b
\tau^\dagger=T^{-1}\circ t^\ddagger \circ T.
\label{new1}
\e

We note that $T$ also maps the first integrals (\ref{si2a}), (\ref{si2b})
of the hierarchy (\ref{mod1b}), with the identifications
(\ref{id2a})---(\ref{id2c}) ($g_n=-2$) to the first integrals
(\ref{si1a}), (\ref{si1b}) of the hierarchy (\ref{mod1a}), with the 
identifications (\ref{id1a})---(\ref{id1c}) ($g_n=2$).

We now consider a corresponding scaling transformation between our
hierarchy (\ref{mod1}) for $g_n=2$, and the same hierarchy for $g_n=-2$.
This scaling transformation is induced from that between the hierarchies
(\ref{mod1a}) and (\ref{mod1b}) as $(U,V,x)=
(\lambda^{-1} \bar U,\lambda^{-1} \bar V,\lambda\xi)$. The hierarchy
(\ref{mod1}), i.e.
\b
\left({\bf F}'[{\bf U}]\right)^{\dagger}{\bf K}_n[{\bf F}[{\bf U}]]
+(c_n,d_n)^T=0,
\label{hie1}
\e
where
\b
{\bf K}_n[{\bf F}[{\bf U}]]={\bf L}_n[{\bf F}[{\bf U}]]
+\sum_{i=1}^{n-1}h_i{\bf L}_i[{\bf F}[{\bf U}]]
+2\left(\begin{array}{c} 0 \\ x \end{array}\right),
\label{bbb}
\e
then becomes (again all derivatives are now with respect to $\xi$)
\b
\left({\bf F}'[{\bf \bar U}]\right)^{\dagger}{\bf K}_n[{\bf F}[{\bf \bar U}]]
+(\bar c_n,\bar d_n)^T=0,
\label{hie2}
\e
with
\b
{\bf K}_n[{\bf F}[{\bf \bar U}]]={\bf L}_n[{\bf F}[{\bf \bar U}]]
+\sum_{i=1}^{n-1}H_i{\bf L}_i[{\bf F}[{\bf \bar U}]]
-2\left(\begin{array}{c} 0 \\ \xi \end{array}\right),
\label{aaa}
\e
and where we have identified $c_n=-\bar c_n$ and $d_n=-\bar d_n$.
That is, we have the BT from (\ref{hie2}), (\ref{aaa}) to (\ref{hie1}),
(\ref{bbb}),
\ba
(U,V,x) & = & (\lambda^{-1} \bar U,\lambda^{-1} \bar V,\lambda\xi), \\
\bar c_n & = & -c_n, \\
\bar d_n & = & -d_n, \\
H_i & = & h_i\lambda^{n-i},
\ea
with $\lambda^{n+1}=-1$,
which, by a convenient abuse of notation, we also refer to as the 
transformation $T$. 

We now consider the composition $T^{-1}\circ \tilde t \circ T$, i.e.\ 
$(\hat\psi,\hat s,\hat l_n,\hat m_n)\rightarrow
(\tilde\phi,\tilde p,\tilde e_n,\tilde f_n)\rightarrow
(\tilde U,\tilde V,\tilde c_n,\tilde d_n)\rightarrow
(\hat U,\hat V,\hat c_n,\hat d_n)$. This BT turns out to be precisely
the BT $\hat\tau$. Thus we have the relation
\b
\hat\tau=T^{-1}\circ \tilde t \circ T.
\label{new2}
\e

We also consider the composition $T^{-1}\circ \hat t \circ T$, i.e.\
$(\bar U,\bar V,\bar c_n,\bar d_n)\rightarrow
(U,V,c_n,d_n)\rightarrow
(\phi,p,e_n,f_n)\rightarrow
(\psi,s,l_n,m_n)$. This BT is $\tilde \tau$, and so we have
\b
\tilde \tau=T^{-1}\circ \hat t \circ T.
\label{new3}
\e

Finally, consideration of the BT $T^{-1}\circ S\circ T$, i.e.\
$(\hat U,\hat V,\hat c_n,\hat d_n)\rightarrow
(\tilde U,\tilde V,\tilde c_n,\tilde d_n)\rightarrow
(U,V,c_n,d_n)\rightarrow
(\bar U,\bar V,\bar c_n,\bar d_n)$, leads to the conclusion
\b
\sigma=T^{-1}\circ S\circ T.
\e

Thus we see that our transformation $T$ is a mapping of Figure Two
into Figure One, but for a different independent variable, $x$ in
Figure One being related to $\xi$ in figure Two by $x=\lambda \xi$
where $\lambda^{n+1}=-1$. Thus of course the relation
$\tau^{\dagger}=\tilde \tau\circ \sigma\circ\hat \tau$ (Figure Two)
is mapped into the relation $t^{\ddagger}=\hat t\circ S\circ\tilde t$
(Figure One).

Our transformation $T$ has some important consequences. It tells us
that the pattern of BTs obtained from our second sequence of Painlev\'e
hierarchies can be related to that obtained from our first sequence of 
Painlev\'e hierarchies. If we had only considered one sequence (e.g.\
the first) it might not have been obvious how to obtain a sequence
(the second) having the pattern of BTs corresponding to $\tau^\dagger$.

The reason why this might not have been obvious is that, for $P_{IV}$,
the BTs ``tilde'' and `hat'' are believed to be independent. This then
leads us on to another of the important consequences of our results:
the BTs ``tilde'' and `hat'' for $P_{IV}$ are not independent, but are
related by a trivial scaling of $P_{IV}$. That is, {\em there is only one 
nontrivial fundamental BT for $P_{IV}$}. This is in contrast to the claim
in \cite{BCH95} that $P_{IV}$ has two nontrivial fundamental BTs (``tilde'' 
and `hat'').

Let us present our results for $P_{IV}$ explicitly. For $n=1$ we may take 
$\lambda=i$ and so our transformation $T$ from (\ref{p4s}) with the 
identification (\ref{id2a})---(\ref{id2c}),
\b
s_{\xi\xi}=\frac{1}{2}\frac{s_\xi^2}{s}+\frac{3}{2}s^3+4\xi s^2+2\left[
\xi^2-a_1\right]s-\frac{1}{2}\frac{b_1^2}{s},
\label{ss}
\e
to (\ref{p4p}) with the identification (\ref{id1a})---(\ref{id1c}),
\b
p_{xx}=\frac{1}{2}\frac{p_x^2}{p}+\frac{3}{2}p^3+4xp^2+2\left[
x^2-A_1\right]p-\frac{1}{2}\frac{B_1^2}{p},
\label{pp}
\e
is
\b
p=is,
\qquad{}
x=i\xi,
\qquad{}
a_1=-A_1
\qquad{}
b_1=-B_1.
\label{T1}
\e
The same transformation $T$ provides a BT from (\ref{p4y}) in $\bar y$ and
$\xi$, with upper choice of sign and the identification
(\ref{id4a})---(\ref{id4c}), 
\b
\bar y_{\xi\xi}=\frac{1}{2}\frac{\bar y_\xi^2}{y}+\frac{3}{2}\bar 
y^3+4\xi\bar y^2
+2\left(\xi^2-\bar \alpha_1\right)\bar y-\frac{1}{2}\frac{\bar \beta_1^2}{\bar y},
\label{qq}
\e
to (\ref{p4y}) in $y$ and $x$, with lower choice 
of sign and the identification (\ref{id3a})---(\ref{id3c}),
\b
y_{xx}=\frac{1}{2}\frac{y_x^2}{y}+\frac{3}{2}y^3+4xy^2+2\left(
x^2-\alpha_1\right)y-\frac{1}{2}\frac{\beta_1^2}{y},
\label{yy}
\e
i.e.
\b
y=i\bar y,
\qquad{}
x=i\xi,
\qquad{}
\bar \alpha_1=-\alpha_1
\qquad{}
\bar\beta_1=-\beta_1.
\label{T2}
\e

In order to show explicitly that $P_{IV}$ has only one fundamental BT
it is enough to show that (\ref{new3}) holds, i.e. that
\b
\tilde \tau = T^{-1}\circ \hat t \circ T.
\e
Here $\hat t$ is the BT (\ref{hatp}), with parameter shifts (\ref{hat3})
and (\ref{hat4}), i.e.
\b
p=-\frac{y_x+y^2+2xy+1+A_1-B_1/2}{2y},
\label{ypbt}
\e
and
\ba
\alpha_1 & = & \frac{1}{4}(2-2A_1-3B_1),  \label{yp1} \\
\beta_1 & = & \frac{1}{2}(2+2A_1-B_1), \label{yp2}
\ea
from (\ref{yy}) to (\ref{pp}).
Meanwhile, $\tilde \tau$ is the BT (\ref{tiltau}), with parameter shifts
(\ref{hat3a}), (\ref{hat4a}), i.e.
\b
s=\frac{\bar y_\xi-\bar y^2-2\xi\bar y+1-a_1+b_1/2}{2\bar y}
\label{qsbt}
\e
\ba
\bar\alpha_1 & = & -\frac{1}{4}(2+2a_1+3b_1),  \label{qs1} \\
\bar\beta_1 & = & -\frac{1}{2}(2-2a_1+b_1), \label{qs2}
\ea
from (\ref{qq}) to (\ref{ss}). It is easy to show that under the
transformation $T$, i.e. when (\ref{T1}) and (\ref{T2}) hold, 
equations (\ref{qsbt})---(\ref{qs2}) are mapped onto equations 
(\ref{ypbt})---(\ref{yp2}). Thus the ``tilde'' and ``hat'' BTs
of $P_{IV}$ are equivalent under a simple scaling transformation,
and we see that $P_{IV}$ has only one nontrivial fundamental auto-BT.

We note that for $P_{IV}$ itself the transformation $T$ can in fact
be found in \cite{Boiti}, and was also known to the authors of 
\cite{BCH95}. However these last failed to recognize that it provides 
a mapping between the ``tilde'' and ``hat'' BTs of $P_{IV}$.

\section{Conclusions}

We have given an improved method of obtaining auto-BTs and special
integrals for hierarchies of ODEs, and have used this to derive 
auto-BTs and special integrals for two fourth Painlev\'e hierarchies.
We have shown how the known pattern of BTs for $P_{IV}$ can be
extended to hierarchies, observing that the BTs required to do this
turn out to be precisely the Miura maps of the DWW hierarchy. Finally,
we have given a mapping between our two sequences of fourth Painlev\'e 
hierarchies which allows us to relate the BTs derived for these two
sequences: in particular, we have derived the result that $P_{IV}$ has 
in fact only one nontrivial fundamental auto-BT.

\section*{Acknowledgements}

The research of PRG and AP is supported in part by the DGESYC under contract
BFM2002-02609, and that of AP by the Junta de Castilla y Le\'on under contract
SA011/04. NJ's research is supported by the Australian Research Council 
Discovery Project Grant \#DP0208430. PRG currently holds a Ram\'on y Cajal
research fellowship awarded by the Ministry of Science and Technology of
Spain, which support is gratefully acknowledged. PRG and AP thank NJ for her 
invitation to visit the University of Sydney in July-August 2002, where this 
work was carried out, and everybody at the School of Mathematics in Sydney for 
their kind hospitality during their stay there.

\newpage

\samepage{
\begin{center}
\begin{tabular}{ccc} \\
$\left(\begin{array}{c}c_n\\d_n\end{array}\right)\sim
\left(\begin{array}{c}\alpha_n\\ \beta_n\end{array}\right)$ & & 
$\left(\begin{array}{c}\tilde c_n\\\tilde d_n\end{array}\right)\sim
\left(\begin{array}{c}\tilde \alpha_n\\\tilde \beta_n\end{array}\right)$
\\ \vspace{-3mm}
& \phantom{\mbox{}\hskip 20mm\mbox{}} & \\ \vspace{-25mm}
$\left(\begin{array}{c}U\\V\end{array}\right)$ &
$\stackrel{S}{\vector(-1,0){130}}$ &
$\left(\begin{array}{c}\tilde U\\ \tilde V\end{array}\right)$ \\
$\begin{array}{cc}
\begin{array}{c} \\ \vspace{40mm} \\ \hat t \end{array}&
{\vector(0,-1){130}}
\end{array}$
& & 
$\begin{array}{cc}
\begin{array}{c}
\vspace{25mm} \\
{\vector(0,1){130}}
\end{array}
& 
\begin{array}{c} \vspace{24mm} \\ \tilde t \end{array}
\end{array}$
\\ \\
$\left(\begin{array}{c}\phi\\p\end{array}\right)$ &
$\stackrel{\vector(-1,0){130}}{t^{\ddagger}}$ &
$\left(\begin{array}{c}\tilde \phi\\ \tilde p\end{array}\right)$ \\
$\left(\begin{array}{c}e_n\\f_n\end{array}\right)\sim
\left(\begin{array}{c}A_n\\B_n\end{array}\right)$
& &
$\left(\begin{array}{c}\tilde e_n\\\tilde f_n\end{array}\right)\sim
\left(\begin{array}{c}\tilde A_n\\\tilde B_n\end{array}\right)$ \\ \\
\end{tabular}
\end{center}
\begin{center}
Figure One: Decomposition of the BT $t^{\ddagger}$ for $P_{IV}$ hierarchies
($g_n=2$).
\end{center}

\begin{center}
\begin{tabular}{ccc} \\
$\left(\begin{array}{c}\bar c_n\\\bar d_n\end{array}\right)\sim
\left(\begin{array}{c}\bar\alpha_n\\ \bar\beta_n\end{array}\right)$ & & 
$\left(\begin{array}{c}\hat c_n\\\hat d_n\end{array}\right)\sim
\left(\begin{array}{c}\hat \alpha_n\\ \hat \beta_n\end{array}\right)$
\\ \vspace{-3mm}
& \phantom{\mbox{}\hskip 20mm\mbox{}} & \\ \vspace{-25mm}
$\left(\begin{array}{c}\bar U\\ \bar V\end{array}\right)$ &
$\stackrel{\sigma}{\vector(-1,0){130}}$ &
$\left(\begin{array}{c}\hat U\\ \hat V\end{array}\right)$ \\
$\begin{array}{cc}
\begin{array}{c} \\ \vspace{40mm} \\ \tilde \tau \end{array}&
{\vector(0,-1){130}}
\end{array}$
& & 
$\begin{array}{cc}
\begin{array}{c}
\vspace{25mm} \\
{\vector(0,1){130}}
\end{array}
& 
\begin{array}{c} \vspace{24mm} \\ \hat \tau \end{array}
\end{array}$
\\ \\
$\left(\begin{array}{c}\psi\\s\end{array}\right)$ &
$\stackrel{\vector(-1,0){130}}{\tau^{\dagger}}$ &
$\left(\begin{array}{c}\hat \psi\\ \hat s\end{array}\right)$ \\
$\left(\begin{array}{c}l_n\\m_n\end{array}\right)\sim
\left(\begin{array}{c}a_n\\b_n\end{array}\right)$
& &
$\left(\begin{array}{c}\hat l_n\\\hat m_n\end{array}\right)\sim
\left(\begin{array}{c}\hat a_n\\\hat b_n\end{array}\right)$ \\ \\
\end{tabular}
\end{center}
\begin{center}
Figure Two: Decomposition of the BT $\tau^{\dagger}$ for $P_{IV}$ hierarchies
($g_n=-2$).
\end{center}
}


\begin{thebibliography}{99}

\bibitem{P00}    P.~Painlev\'e, {\em Bull. Soc. Math. Fr.} {\bf 28} 201--261 
                 (1900).
\bibitem{P02}    P.~Painlev\'e, {\em Acta Math.} {\bf 25} 1--85 (1902).
\bibitem{G10}    B.~Gambier, {\em Acta Math.} {\bf 33} 1--55 (1910).
\bibitem{Ince}   E.~L.~Ince, ``Ordinary Differential Equations,'' (Dover, New
                 York, 1956).
\bibitem{C11}    J.~Chazy, {\em Acta Math.} {\bf 34} 317-385 (1911).
\bibitem{G12}    R.~Garnier, {\em Ann. Sci. \'Ecole Normale Sup.} {\bf 48}
                 1-126 (1912).
\bibitem{E71}    H.~Exton, {\em Rend. Mat. (6)} {\bf 6} 419-462 (1973).
\bibitem{M85a}   I.~P.~Martynov, {\em Differents. Uravn.} {\bf 21} 764-771
                 (1985).
\bibitem{M85b}   I.~P.~Martynov, {\em Differents. Uravn.} {\bf 21} 037-946
                 (1985).
\bibitem{B64}    F.~J.~Bureau, {\em Ann. Mat. Pura Appl. (IV)} {\bf 66}
                 1-116 (1964).
\bibitem{AS77}   M.~J.~Ablowitz and H.~Segur, {\em Phys. Rev. Lett.} {\bf 38}
                 1103--1106 (1977).
\bibitem{K89a}   S.~Kowalevski, {\em Acta Math.} {\bf 12} 177-232 (1889).
\bibitem{K89b}   S.~Kowalevski, {\em Acta Math.} {\bf 14} 81-93 (1889).
\bibitem{Air}    H.~Airault, {\em Stud. Appl. Math.} {\bf 61} 31-53 (1979).
\bibitem{FA82}   A.~S.~Fokas and M.~J.~Ablowitz {\em J. Math. Phys.} {\bf 23}
                 2033--2042 (1982).
\bibitem{GLS}    V.~Gromak, I.~Laine and S.~Shimomura, ``Painlev\'e
                 Differential Equations in the Complex Plane,'' (de Gruyter,
                 Berlin, 2002).
\bibitem{MJ99}   U.~Mu\u gan and F.~Jrad, {\em J. Phys. A} {\bf 32}  
                 7933--7952 (1999).
\bibitem{MJ02}   U.~Mu\u gan and F.~Jrad, {\em J. Nonlinear Math. Phys.}
                 {\bf 9} 282--310 (2002).
\bibitem{MJp}    U.~Mu\u gan and F.~Jrad, 
                 {\em Z. Naturforsch. A} {\bf 59} 163-180 (2004).
\bibitem{C00}    C.~M.~Cosgrove, {\em Stud. Appl. Math.} {\bf 104} 1-65 
                 (2000).
\bibitem{GP99a}  P.~R.~Gordoa and A.~Pickering, {\em Europhys. Lett.} {\bf 47} 
                 21-24 (1999). 
\bibitem{GP99b}  P.~R.~Gordoa and A.~Pickering, {\em J. Math. Phys.} {\bf 40}
                 5749-5786 (1999). 
\bibitem{GP00}   P.~R.~Gordoa and A.~Pickering, {\em J. Phys. A} {\bf 33} 557-567 
                 (2000). 
\bibitem{GJP01}  P.~R.~Gordoa, N.~Joshi and A.~Pickering, {\em Publ. Res. Inst.
                 Math. Sci. (Kyoto)} {\bf 37} 327--347 (2001).
\bibitem{B75}    L.~J.~F.~Broer, {\em Appl. Sci. Res.} {\bf 31} 377-395 (1975).
\bibitem{K75a}   D.~J.~Kaup, 
                 {\em Prog. Theor. Phys.} {\bf 54} 72-78 (1975).
\bibitem{K75b}   D.~J.~Kaup, 
                 {\em Prog. Theor. Phys.} {\bf 54} 396-408 (1975).
\bibitem{JM}     M.~Jaulent and J.~Miodek, {\em Lett. Math. Phys.} {\bf 1}
                 243-250 (1976).
\bibitem{MY}     V.~B.~Matveev and M.~I.~Yavor, {\em Annales de l'Institut
                 Henri Poincar\'e} {\bf 31} 25-41 (1979).
\bibitem{LMA80}  L.~Mart\'{\i}nez Alonso, {\em J. Math. Phys.} {\bf 21} 2342-2349 
                 (1980).
\bibitem{Kup}    B.~A.~Kupershmidt, {\em Commun. Math. Phys.}
                 {\bf 99} 51-73 (1985).
\bibitem{sachs}  R.~L.~Sachs, {\em Physica D} {\bf 30} 1-27 (1988).
\bibitem{GJP02}  P.~R.~Gordoa, N.~Joshi and A.~Pickering, {\em Publ.\ Res.\ 
                 Inst.\ Math.\ Sci.\ (Kyoto)} {\bf 39} 435-449 (2003).
\bibitem{AP02}   A.~Pickering, {\em Teoret.\ i Matem. Fizika} {\bf 137} 445-456 
                 (2003); {\em Theoret.\ and Math.\ Phys.} {\bf 137} 1733-1742 
                 (2003).
\bibitem{Lax}    P.~D.~Lax, {\em SIAM Rev.} {\bf 18} 351-375 (1976).
\bibitem{W83}    J.~Weiss, {\em J. Math. Phys.} {\bf 24} 1405-1413 (1983).
\bibitem{CJP99}  P.~A.~Clarkson, N.~Joshi and A.~Pickering, {\em Inverse
                 Problems} {\bf 15} 175-187 (1999).
\bibitem{BCH95}  A.~Bassom, P.~A.~Clarkson and A.~C.~Hicks, {\em Stud. Appl.
                 Math.} {\bf 95} 1-71 (1995).
\bibitem{GJP99}  P.~R.~Gordoa, N.~Joshi and A.~Pickering, {\em Nonlinearity}
                 {\bf 12} 955-968 (1999).
\bibitem{Boiti}  M.~Boiti and F.~Pempinelli, {\em Nuovo Cimento} {\bf 59B}
                 40-58 (1980).

\end{thebibliography}
\end{document}